\documentclass[3p,times,twocolumn]{elsarticle}
 \biboptions{comma,sort&compress}
 
\usepackage{graphicx}
\usepackage{amsfonts, amsmath, amsthm, amssymb}
\usepackage{graphicx}
\usepackage[labelformat=empty]{subfig}
\usepackage{here}
\usepackage{ulem}

\usepackage{xcolor}
\usepackage{ecrc}

\volume{00}

\firstpage{1}

\journalname{Nuclear and Particle Physics Proceedings}

\runauth{}

\jid{nppp}

\jnltitlelogo{Nuclear and Particle Physics Proceedings}

\usepackage{amssymb}

\usepackage[figuresright]{rotating}

\newcommand{\dfourx}{\mathrm{d}^4 x}

\newcommand{\dt}{\mathrm{d}t}
\renewcommand{\Im}{\mathrm{Im}}
\newcommand{\as}{\alpha_s}
\newcommand{\borel}{\hat{\mathcal{B}}}
\newcommand{\quarkthreed}{\langle\overline{q}q\rangle}
\newcommand{\gluonfourd}{\langle\alpha G^2\rangle}
\newcommand{\mixed}{\langle g\overline{q} \sigma G q\rangle}
\newcommand{\gluonsixd}{\langle g^3 G^3\rangle}
\newcommand{\vev}[1]{\langle\Omega|#1|\Omega\rangle}
\newcommand{\angled}[1]{\langle #1\rangle}

\newcommand{\lsr}{\mathcal{R}}
\newcommand{\gev}{\ensuremath{\text{GeV}}}
\newcommand{\mev}{\ensuremath{\text{MeV}}}

\begin{document}

\begin{frontmatter}

\title{ Ground State Mass Predictions of Heavy-Light Hybrids from QCD Sum-Rule Analysis ($J^P=\left\{0^{\pm},\,1^{\pm}\right\}$)
 $^*$}
 \cortext[cor0]{Talk given at 18th International Conference in Quantum Chromodynamics (QCD 17),  3 july - 8 july 2017, Montpellier - FR}
 \author[label1]{J.~Ho\fnref{fn1}\corref{cor1}}
 \cortext[cor1]{PhD student.}
 \fntext[fn1]{Speaker, Corresponding author.}
\ead{j.ho@usask.ca}
\address[label1]{Department of Physics and Engineering Physics, University of Saskatchewan, Saskatoon, SK, S7N 5E2, Canada}
\address[label2]{Department of Physics, University of the Fraser Valley, Abbotsford, BC, V2S 7M8, Canada}
 \author[label2]{D.~Harnett}
\ead{derek.harnett@ufv.ca}
 \author[label1]{T.G.~Steele}
    \ead{tom.steele@usask.ca}

\pagestyle{myheadings}
\markright{ }
\begin{abstract}
We present QCD Laplace sum-rule predictions of ground state masses of heavy-light open-flavour hybrid mesons. Having computed leading-order diagonal correlation functions, including up to dimension six gluon condensate contributions, we extract hybrid mass predictions for all $J^P \in \{0^\pm, 1^\pm\}$, and explore possible mixing effects with conventional meson states. Similarities are found in the mass hierarchy in both charm and bottom systems with some exceptions that are discussed.
\end{abstract}

\end{frontmatter}

\section{Correlation Functions of Heavy-Light Open-Flavour Hybrid Mesons}
The first investigations of heavy-light hybrids using QCD sum-rules were performed by 
Govaerts, Reinders, and Weyers~\cite{GovaertsReindersWeyers1985} (abbreviated GRW).
In that work, they considered four currents covering $J\in\{0,\,1\}$
in an effort to compute a comprehensive collection of ground state hybrid masses.
For each heavy-light hybrid ground state, the square of the predicted mass was
close to the continuum threshold (with separations
of roughly 10--15~$\mev$); GRW noted that even a modest hadronic resonance width would result in concern of contamination from the continuum \cite{GovaertsReindersWeyers1985}.

We review work done in \cite{HoHarnettSteele2016} where we extended the work of GRW \cite{GovaertsReindersWeyers1985} by  
including higher dimensional condensate contributions (dimension-five mixed and dimension-six gluon) in
the calculation of our correlator.
In the case of heavy-light hybrids, condensates involving light quarks
are enhanced by a heavy quark mass allowing for the possibility
of significant contributions to the correlator and to the sum-rules. This was noted previously by GRW \cite{GovaertsReindersWeyers1985}. 
Thus, the dimension-five mixed condensate could become a significant component 
of a QCD sum-rules application to the heavy-light hybrid systems.
Further, sum-rules analyses of closed-flavor, 
heavy hybrid mesons~\cite{ChenKleivSteeleEtAl2013} have demonstrated that the dimension-six gluon condensate can have a significant stabilizing
effect on what were in previous studies ~\cite{GovaertsReindersWeyers1985,GovaertsReindersRubinsteinEtAl1985,GovaertsReindersFranckenEtAl1987} unstable analyses.

We define our open-flavour hybrid interpolating currents in the same fashion as GRW, 
\begin{equation}
j_{\mu}=\frac{g_s}{2} ~ \overline{Q} ~ \Gamma^{\rho} \lambda^a q~  \mathcal{G}^a_{\mu\rho},
\end{equation}
where $g_s$ is the strong coupling and $\lambda^a$ are the Gell-Mann matrices.
The field $Q$ represents a heavy charm or bottom quark with mass $M_Q$
while $q$ represents a light up, down, or strange quark with mass $m_q$. 
The Dirac matrix $\Gamma^{\rho}$ satisfies $\Gamma^{\rho} \in \{\gamma^{\rho},\,\gamma^{\rho}\gamma_5\}$
and the tensor $\mathcal{G}^a_{\mu\rho}$, the portion of $j_{\mu}$ containing 
the gluonic degrees of freedom, satisfies 
\begin{equation}
\mathcal{G}^a_{\mu\rho} \in \{G^a_{\mu\rho},\,\tilde{G}^a_{\mu\rho} = \frac{1}{2}\epsilon_{\mu\rho\nu\sigma}G^a_{\nu\sigma}\},
\end{equation}
where $G^a_{\mu\rho}$ is the gluon field strength and $\tilde{G}^a_{\mu\rho}$ is its dual defined using the Levi-Civita 
symbol $\epsilon_{\mu\rho\nu\sigma}$.

For each of the currents defined through $j_{\mu}$ above,
we consider a diagonal, two-point correlation function
\begin{align}
\label{correlator}
  \Pi_{\mu\nu}(q) = & i\int\dfourx e^{i q\cdot x} 
    \vev{\tau j_{\mu}(x)j^{\dag}_{\nu}(0)}\nonumber \\
  = & \frac{q_{\mu}q_{\nu}}{q^2}\Pi^{(0)}(q^2) 
   + \left(\frac{q_{\mu}q_{\nu}}{q^2}-g_{\mu\nu}\right)\Pi^{(1)}(q^2),
\end{align}
where $\Pi^{(0)}$ probes spin-0 states and $\Pi^{(1)}$ probes spin-1 states.
We will reference each of the $\Pi^{(0)}$ and $\Pi^{(1)}$
according to the $J^{PC}$ combination it would carry in the flavour-symmetric limit; however, to emphasize that the $C$-value can not be taken literally for open-flavour systems, we will enclose it in parentheses.
\begin{center}
\captionof{table}{The $J^{P(C)}$ combinations probed through different choices of $\Gamma^{\rho}$ and $\mathcal{G}^a_{\mu\rho}$.}
\begin{tabular}[h]{c|c||c}
  $\Gamma^{\rho}$ & $\mathcal{G}^a_{\mu\rho}$ & $J^{P(C)}$\\
  \hline
  $\gamma^{\rho}$ & $G^a_{\mu\rho}$ & $0^{+(+)},\,1^{-(+)}$ \\
  $\gamma^{\rho}$ & $\tilde{G}^a_{\mu\rho}$ & $0^{-(+)},\,1^{+(+)}$ \\
  $\gamma^{\rho}\gamma_5$ & $G^a_{\mu\rho}$ & $0^{-(-)},\,1^{+(-)}$ \\
  $\gamma^{\rho}\gamma_5$ & $\tilde{G}^a_{\mu\rho}$ & $0^{+(-)},\,1^{-(-)}$
\label{JPC_table}
\end{tabular}
\end{center}

We utilize the operator product expansion (OPE) to calculate the correlators~(\ref{correlator});
within the OPE, perturbation theory is supplemented by non-perturbative terms,
each of which is the product of a perturbatively computed Wilson coefficient 
and a non-zero vacuum expectation value (VEV), also referred to as a condensate.  
We include
\begin{gather}
  \quarkthreed=\angled{\overline{q}_i^{\alpha} q_i^{\alpha}}
    \label{condensate_quark_three}\\
  \gluonfourd=\angled{\alpha_s G^a_{\mu\nu} G^a_{\mu\nu}}\label{condensate_gluon_four}\\
  \mixed=\angled{g_s \overline{q}_i^{\alpha}\sigma^{\mu\nu}_{ij}
    \lambda^a_{\alpha\beta} G^a_{\mu\nu} q_j^{\beta}}
    \label{condensate_mixed}\\
  \gluonsixd=\angled{g_s^3 f^{abc} G^a_{\mu\nu}G^b_{\nu\rho}G^c_{\rho\mu}},\label{condensate_gluon_six}
\end{gather}
where the VEVs~(\ref{condensate_quark_three})--(\ref{condensate_gluon_six}) 
are respectively referred to as
the 3d~quark condensate (i.e. the dimension-three quark condensate),
the 4d~gluon condensate,
the 5d~mixed condensate, and
the 6d~gluon condensate.

\begin{figure}[ht!]
\subfloat[]{%
    \includegraphics[width=.142\textwidth]{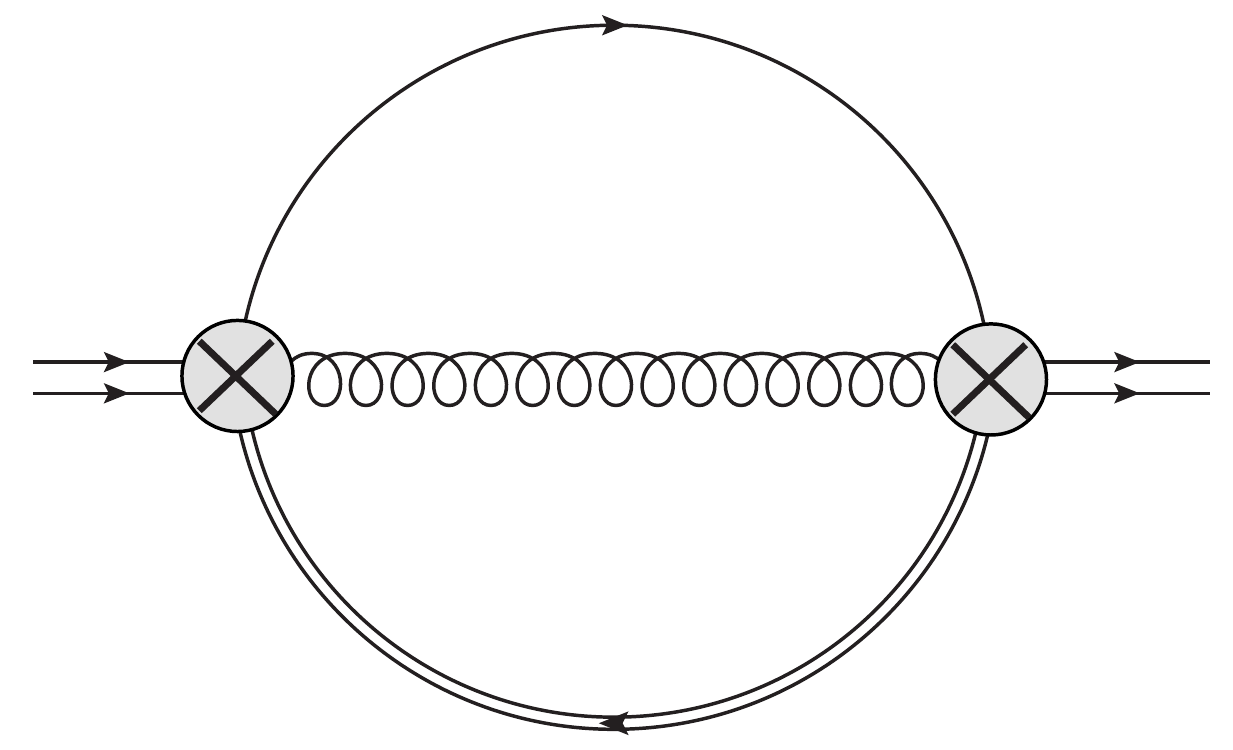}}\hfill
  \subfloat[]{%
    \includegraphics[width=.142\textwidth]{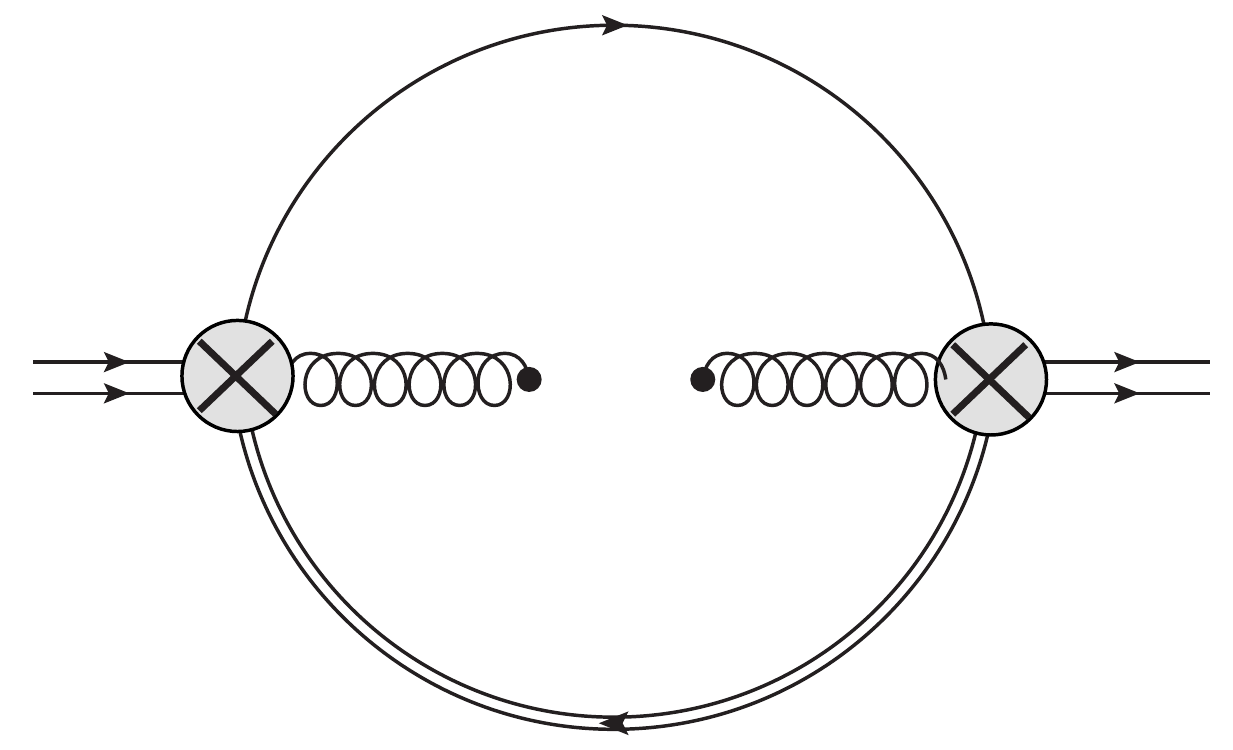}}\hfill
  \subfloat[]{%
    \includegraphics[width=.142\textwidth]{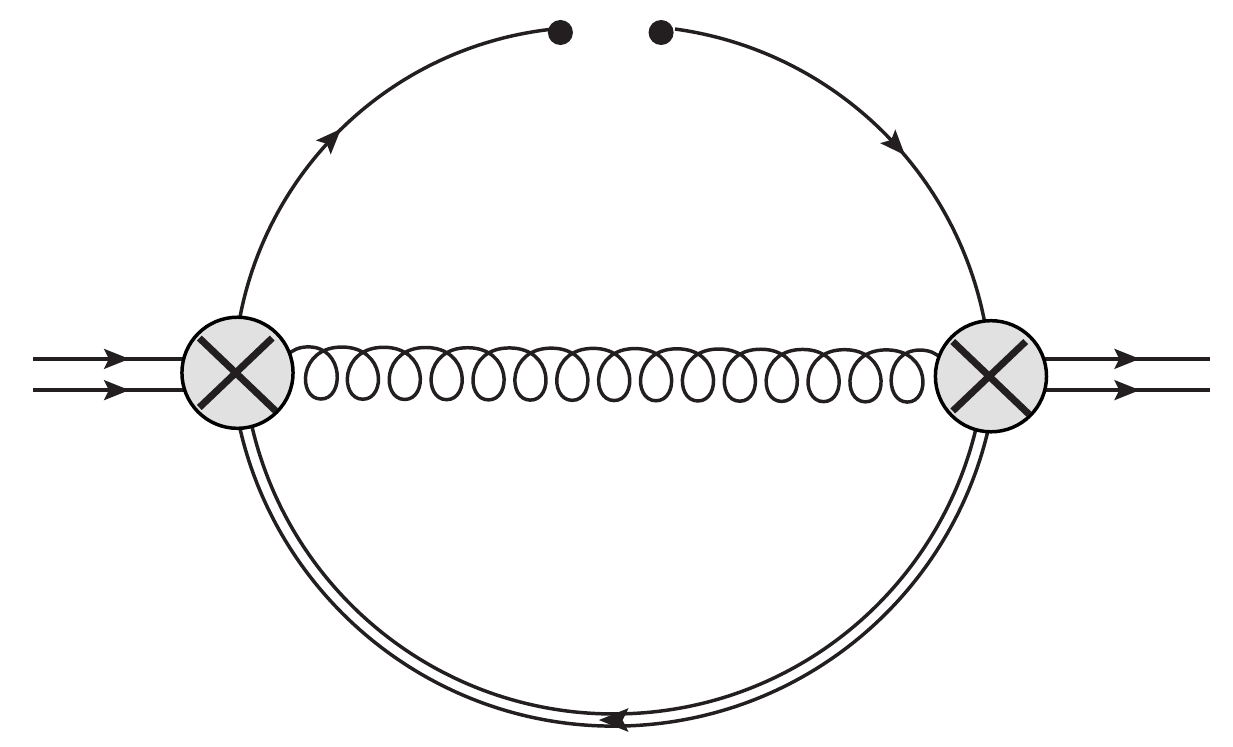}}\hfill
  \subfloat[]{%
  	\includegraphics[width=.142\textwidth]{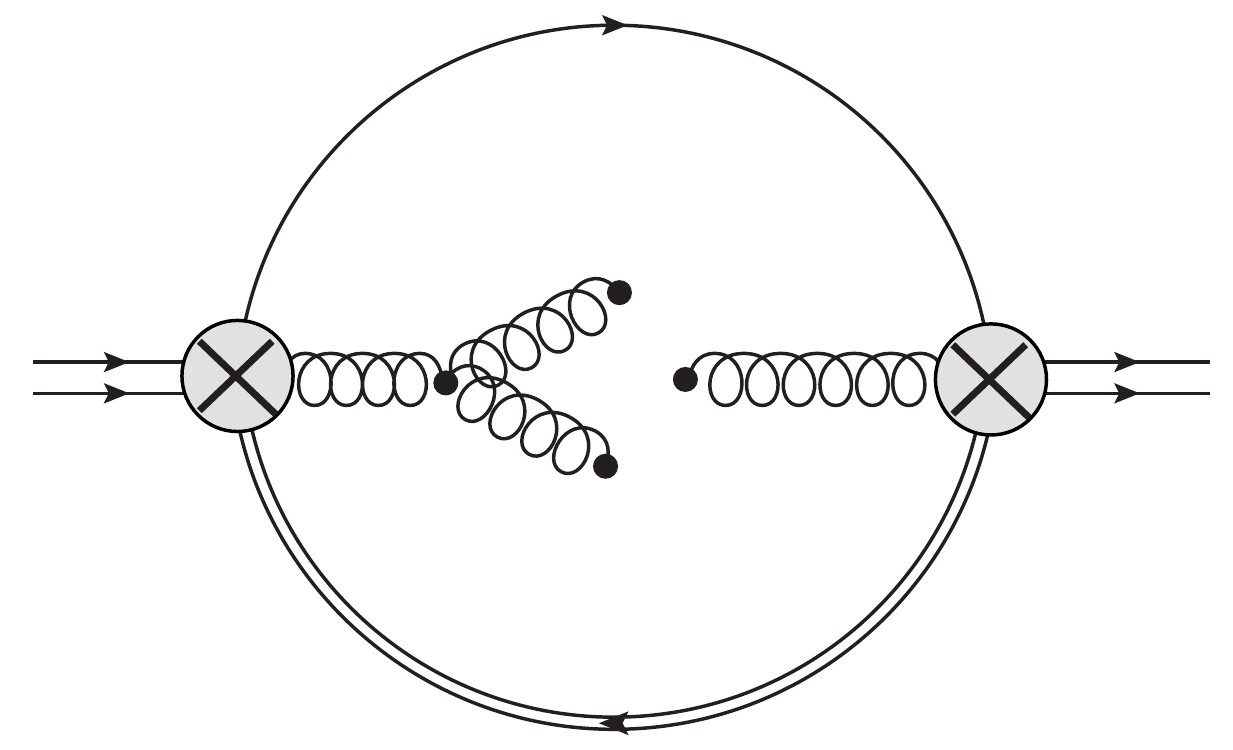}}\hfill
  \subfloat[]{%
    \includegraphics[width=.142\textwidth]{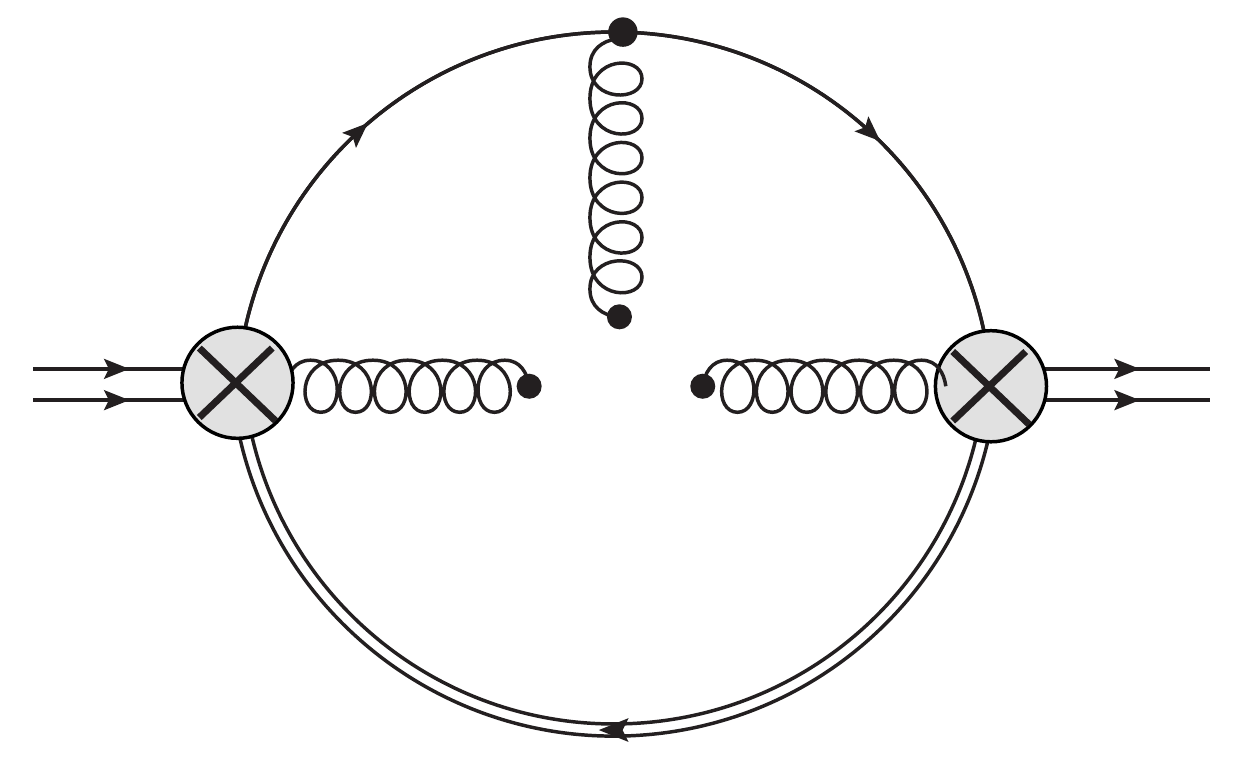}}\hfill
  \subfloat[]{%
  	\includegraphics[width=.142\textwidth]{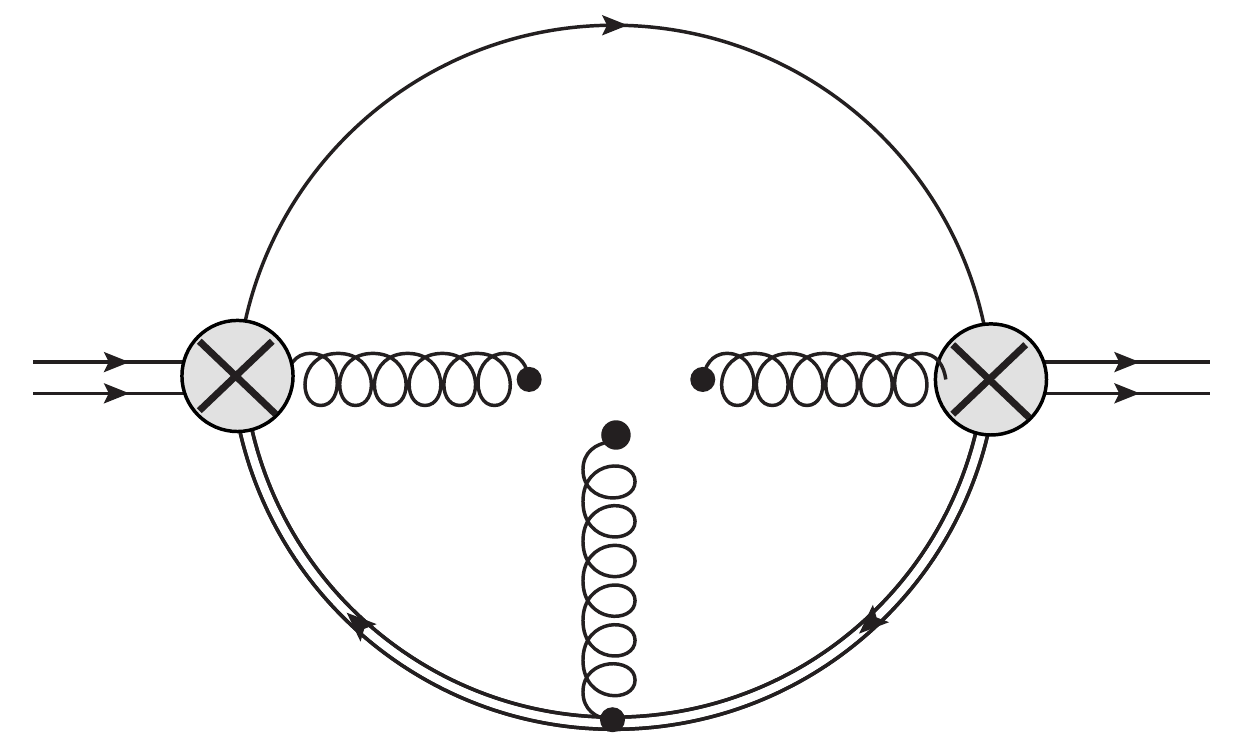}}\hfill
  \subfloat[]{%
    \includegraphics[width=.142\textwidth]{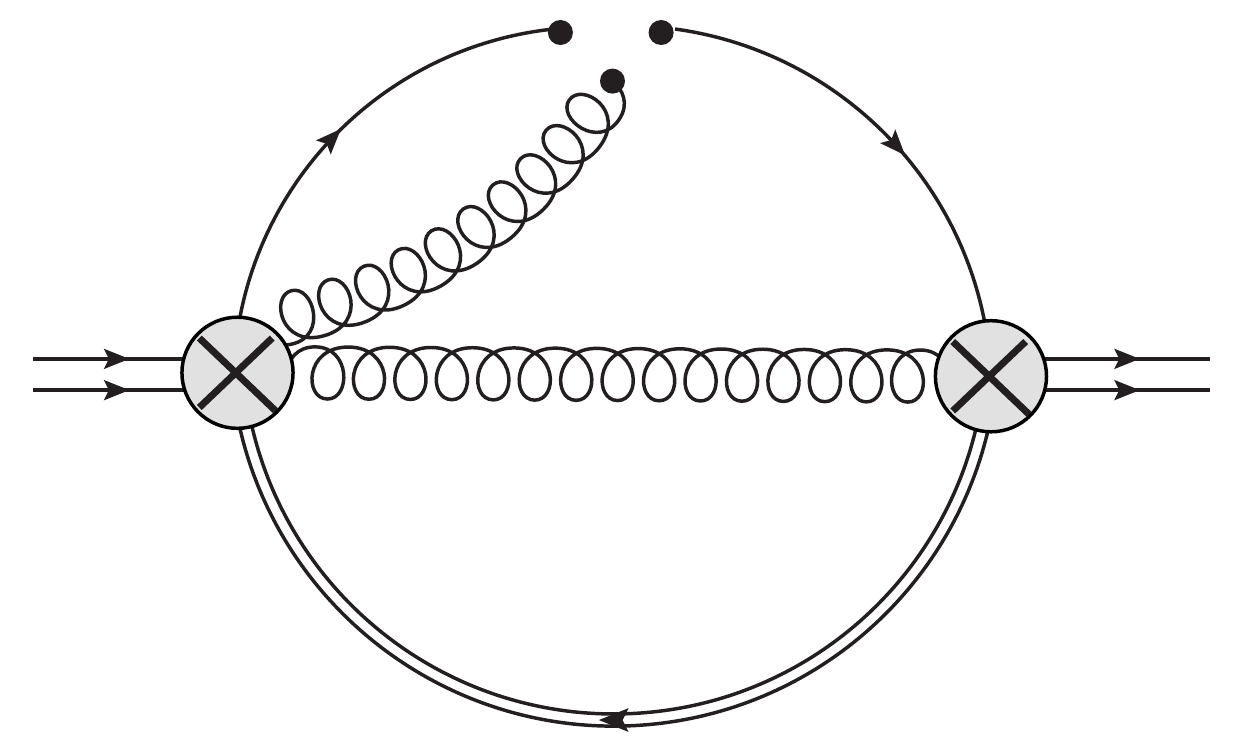}}\hfill
  \subfloat[]{%
    \includegraphics[width=.142\textwidth]{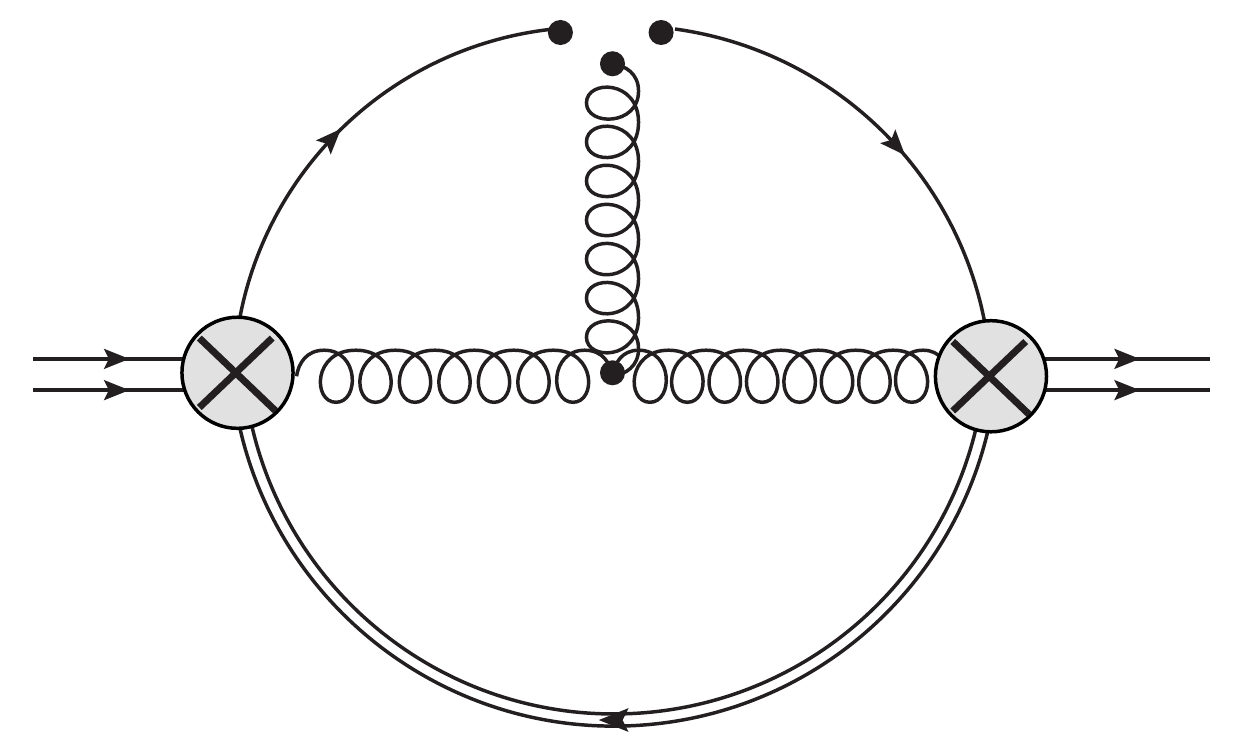}}\hfill
  \subfloat[]{%
    \includegraphics[width=.142\textwidth]{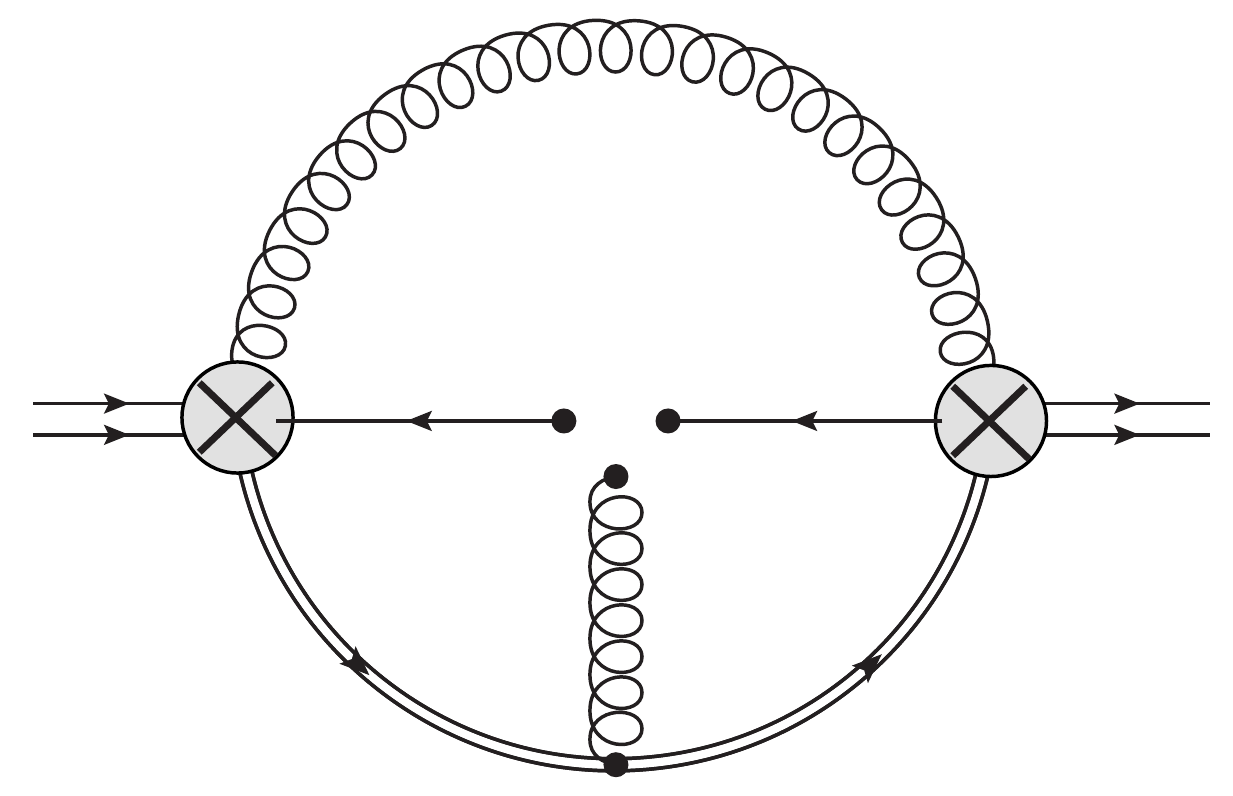}}\hfill
  \subfloat[]{%
    \includegraphics[width=.142\textwidth]{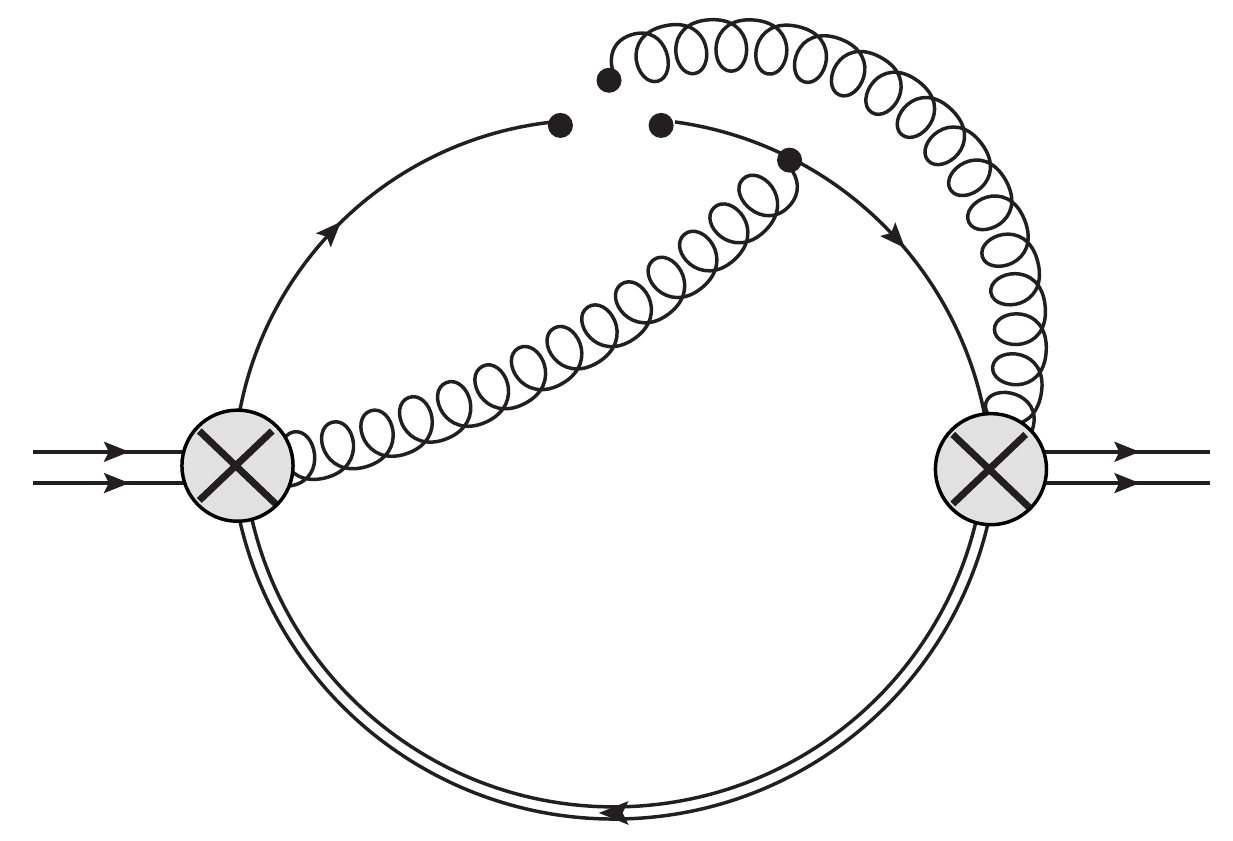}}\hfill
  \subfloat[]{%
    \includegraphics[width=.142\textwidth]{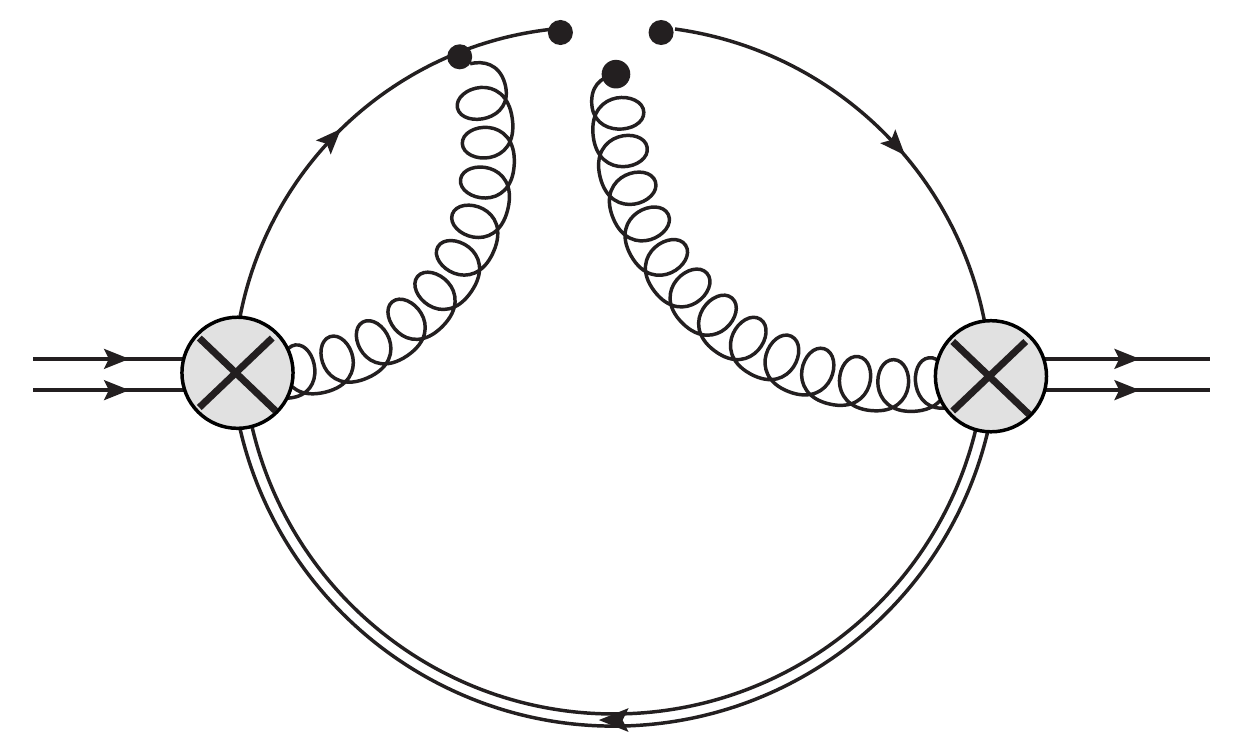}}\hfill
  \subfloat[]{%
    \includegraphics[width=.142\textwidth]{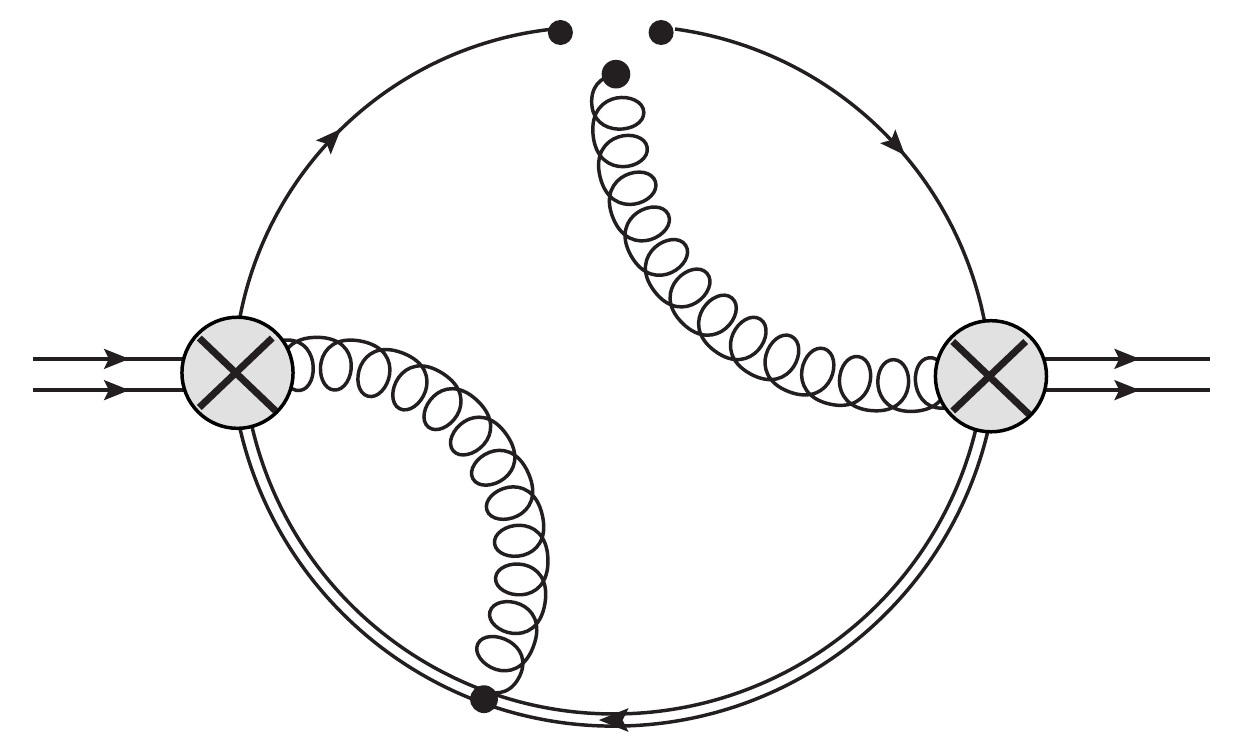}}\hfill
  \subfloat[]{\hspace{1.25cm}
    \includegraphics[width=.142\textwidth]{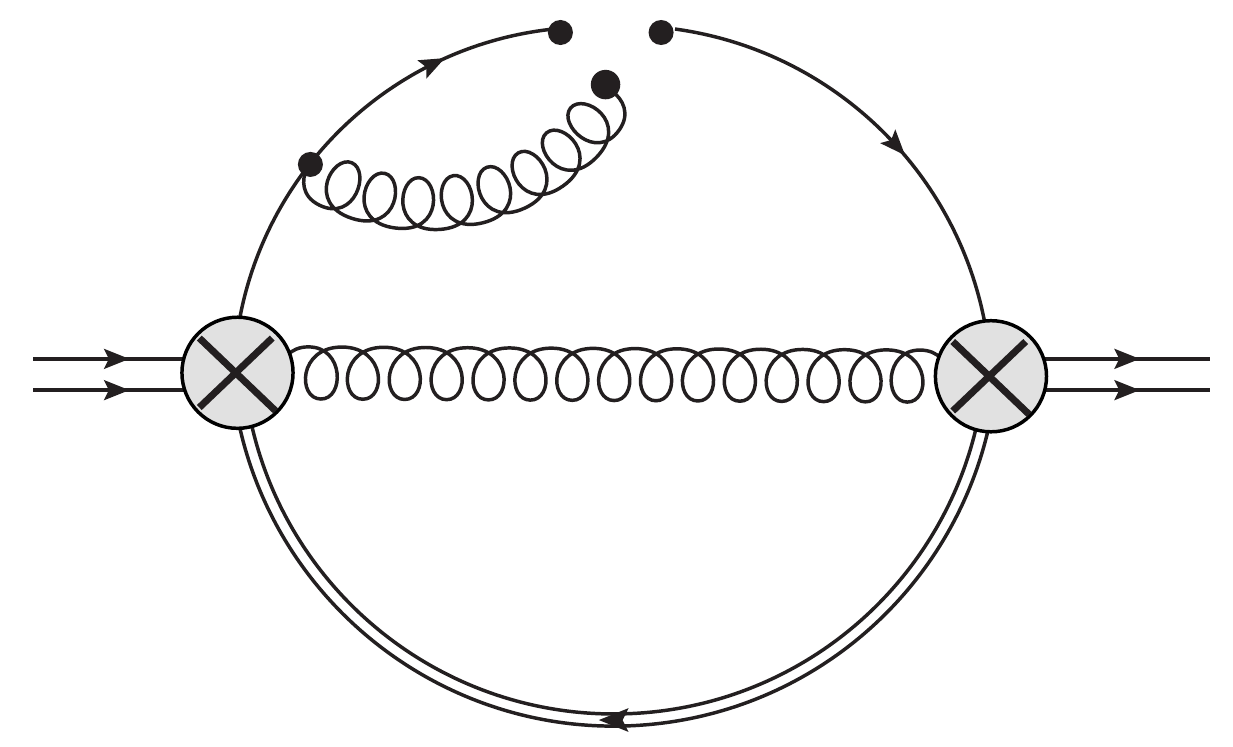}}\hfill
  \subfloat[]{%
    \includegraphics[width=.142\textwidth]{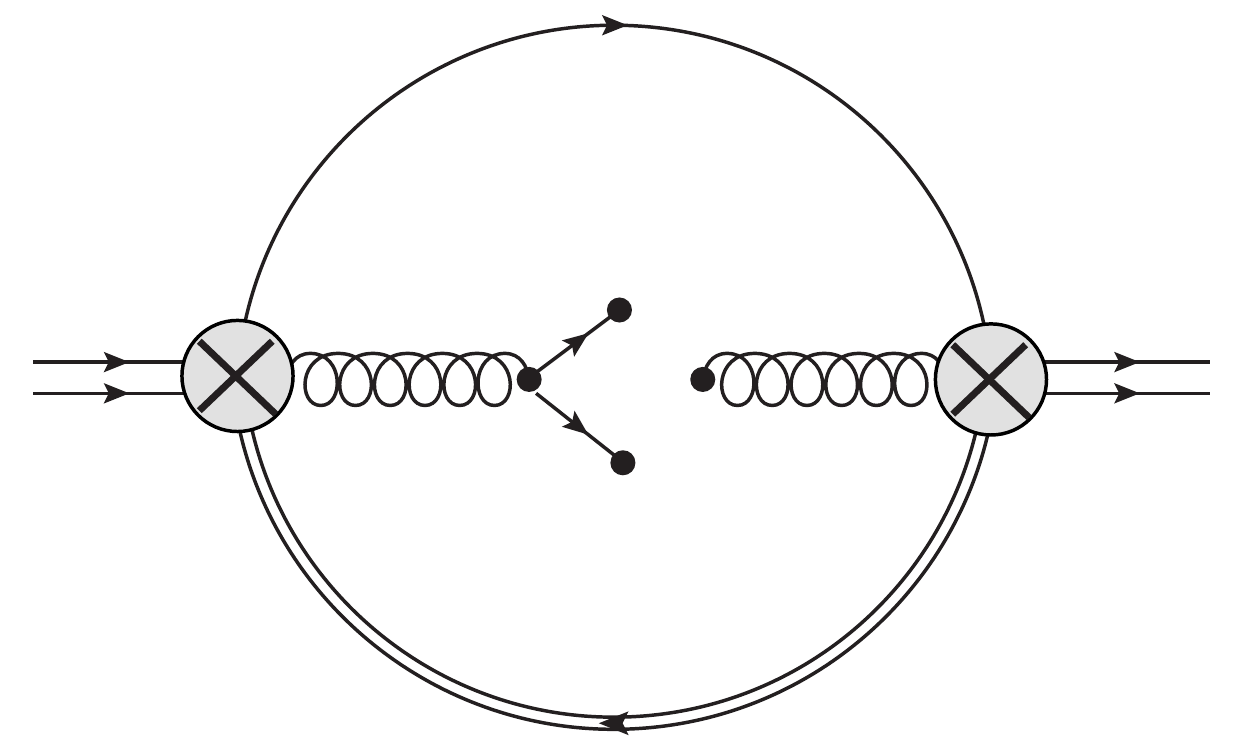}\hspace{1.25cm}}\hfill
\caption{The Feynman diagrams calculated for the correlator~(\ref{correlator}). Single solid lines correspond to light quark propagators whereas double solid lines correspond to heavy quark propagators. All Feynman diagrams are drawn using JaxoDraw~\protect\cite{BinosiTheussl2004}}
\label{feynman_diagrams}
\end{figure}

The Wilson coefficients are 
computed to leading-order (LO) in $g_s$
(see \cite{BaganAhmadyEliasEtAl1994} for a review of calculational methods).
Light quark masses are included in perturbation theory through a light
quark mass expansion, but have been set to zero in all other OPE terms.
The contributing Feynman diagrams are depicted in 
Figure~\ref{feynman_diagrams}.
Dimensional regularization in $D=4+2\epsilon$ spacetime dimensions at renormalization scale $\mu^2$ is employed for divergent integrals.
We use the program TARCER~\cite{MertigScharf1998}
to reduce two-loop integrals to a small collection of 
simpler integrals, all of which are well-known for the diagrams under consideration.
The results of calculations are given in Ref.~\cite{HoHarnettSteele2016} and are omitted here for brevity.

We briefly review the details of the Laplace sum-rule (LSR) methodology~\cite{ShifmanVainshteinZakharov1979}.
The formation of the sum-rule begins with a dispersion relation phrased in terms of Euclideanized momentum $Q^2=-q^2$, 
\begin{equation}\label{dispersion_relation}
  \Pi(Q^2)=\frac{Q^8}{\pi}\int_{t_0}^{\infty}
  \frac{\Im\Pi(t)}{t^4(t+Q^2)}
  \,\dt+\cdots,\ Q^2>0
\end{equation}
where $\cdots$ represents subtraction constants (a polynomial in $Q^2$),
and $t_0$ represents the appropriate physical threshold. The dispersion relation  (\ref{dispersion_relation}), satisfied by each of the components $\{\Pi^{(0)},\,\Pi^{(1)}\}$, 
encapsulates quark-hadron duality: a connection between hadrons and QCD. The left side of (\ref{dispersion_relation}) is calculated from the OPE discussed earlier in this section, while $\Im\Pi(t)$ on the right side represents the hadronic spectral function. In a traditional sum-rules analysis, this spectral function is parameterized in terms of hadronic quantities within a model that are then fit to the OPE result. 
To form our sum-rule, we apply to (\ref{dispersion_relation}) the Borel transform,
\begin{equation}\label{borel}
  \borel =\lim_{\stackrel{N,Q^2\rightarrow\infty}{\tau=N/Q^2}}
  \frac{(-Q^2)^N}{\Gamma(N)}\left(\frac{d}{dQ^2}\right)^N\,.
\end{equation}
This enhances the lower-lying ground state resonances and suppresses excited states described by our correlator, and conveniently eliminates the subtraction constants in (\ref{dispersion_relation}). This gives the $k^{\text{th}}$-order Laplace sum-rule 
\begin{align}
\label{lsr_main}
  \lsr_k(\tau) =& \int_{t_0}^{\infty}t^k e^{-t\tau}\frac{1}{\pi}\Im\Pi(t)\,\dt\\
  =&\frac{1}{\tau}\borel\left\{(-Q^2)^k\Pi(Q^2)\right\}.\label{lsr_borel}
\end{align}
Employing a ``single narrow resonance plus continuum'' model for our spectral function
\cite{ShifmanVainshteinZakharov1979}
\begin{equation}\label{single_narrow_resonance}
  \frac{1}{\pi}\Im\Pi(t) = f_H^2 m_{H}^{8} \delta(t-m_H^2)
  +\theta(t-s_0)\frac{1}{\pi}\Im\Pi^{\text{OPE}}(t)
\end{equation}
gives us a final form for our (continuum-subtracted) sum-rule, 
\begin{align}\label{lsr_subtracted}
  \lsr_k(\tau,\,s_0)=&\lsr_k(\tau)
  -\int_{s_0}^{\infty}t^k e^{-t\tau}\frac{1}{\pi}\Im\Pi^{\text{OPE}}(t)\dt\\
  =& f_H^2 m_H^{8+2k} e^{-m_H^2 \tau}
\end{align}
where $m_H$ is the ground state resonance mass, $f_H$ is its coupling strength,
$\theta$ is a Heaviside step function, $s_0$ is the continuum threshold
and $\Im\Pi^{\text{OPE}}$ is the imaginary part of 
the QCD expression for $\Pi$. This form allows us to extract the hadronic mass as a ratio of these continuum-subtracted sum-rules,
\begin{equation}\label{lsr_master}
  \frac{\lsr_1(\tau,\,s_0 )}{\lsr_0(\tau,\,s_0 )}=m_H^2.
\end{equation}
To compute stable mass predictions using (\ref{lsr_master}), we require a suitable range of values for our Borel scale ($\tau$) within which reliable results can be extracted (known as a Borel window). Within this Borel window, we perform a fitting procedure in order to find an optimized value of the continuum onset ($s_0$) associated with our resulting mass prediction. We determine our Borel window by requiring convergence of the OPE, and that the pole mass contributes a certain percentage to the overall mass prediction. We follow our previous work done in closed-flavour heavy hybrid systems \cite{ChenKleivSteeleEtAl2013}; to enforce OPE convergence and obtain an upper-bound on the window, we require that contributions to the 4d condensate be less than one-third that of the perturbative contribution, and the 6d gluon condensate contribute less than one-third of the 4d condensate contributions. For a lower-bound on the Borel window, we require a pole contribution of at least $10\%$.
\section{Mixing Effects}
A consequence of the open-flavour systems in question is the preclusion of exotic quantum numbers; as no definite C-parity number exists for the $J^P$ systems in question, we cannot access clean, ``smoking gun'' signals that exotic quantum numbers might allow. As such, we must consider possible effects of mixing with conventional meson systems. For a preliminary examination of these effects, we extend our single narrow resonance model to a double narrow resonance model coupling to a conventional meson state, 
\begin{align}\label{double_narrow_resonance}
  \frac{1}{\pi}\Im\Pi(t) = & f_H^2 m_{H}^{8} \delta(t-m_H^2) + f_{conv}^2 m_{conv}^{8} \delta(t-m_{conv}^2)\nonumber\\
  &+\theta(t-s_0)\frac{1}{\pi}\Im\Pi^{\text{OPE}}(t).
\end{align}
With this change in resonance model, our resulting sum-rule in (\ref{lsr_master}) becomes
\begin{equation}\label{lsr_master_mixed}
\frac{\lsr_1(\tau,\,s_0 )-f_{conv}^2 m_{conv}^{10} e^{-m_{conv}^2 \tau}}{\lsr_0(\tau,\,s_0 )-f_{conv}^2 m_{conv}^{8} e^{-m_{conv}^2 \tau}} = m_H^2.
\end{equation}
While this is no replacement for a full mixing analysis involving off-diagonal contributions, it serves as a simple model to examine possible mixing effects. To examine the effects of conventional state mixing, we have anchored our double narrow resonance model to the corresponding lowest-lying resonance for a given set of quantum numbers using a conventional meson mass $m_{conv}$ obtained from the PDG \cite{OliveEtAl2014}. Along with our mass predictions from the single narrow resonance model, we also report the maximal mixing effect from considering (\ref{lsr_master_mixed}) in Tables \ref{D0_results_table}-\ref{Bs_results_table}; $\delta m_{mix}$ is the increased mass range with mixing up to $\left\vert\frac{f_{conv}}{f_H}\right\vert = \frac{1}{2}$ due to coupling to the lowest-lying conventional state with appropriate quantum numbers according to PDG values summarized in Table \ref{mixing_error}. In all cases, the results of the mixing tends to increase the resulting mass prediction, though the charm flavoured channels appear particularly insensitive to this mixing model.

The quark masses and $\as$ reference values used in our analysis are 
\begin{gather}
 M_c=(1.275 \pm 0.025)\ \gev\\
 M_b=(4.18 \pm 0.03)\ \gev\\
  m_n(2~\gev)=(3.40 \pm 0.25)\ \mev\\
  m_s(2~\gev)=(93.5 \pm 2.5)\ \mev\\
 \as(M_{\tau})=0.330\pm0.014\\
 \as(M_Z)=0.1185\pm0.0006.
\end{gather}
as well as the mass ratios
\begin{align}
  \frac{M_c}{m_{n}} &= 322.6\pm 13.6, & \frac{M_c}{m_{s}} &= 11.73 \pm 0.25,\\
  \frac{M_b}{m_{n}} &= 1460.7\pm 64.0, & \frac{M_b}{m_{s}} &= 52.55 \pm 1.30.
\end{align}
The condensate values used are
\begin{gather}
  \gluonfourd = (0.075 \pm 0.020)\ \gev^4\label{gluonfourdvalue}\\
  \gluonsixd = \left((8.2 \pm 1.0)\ \gev^2\right)\gluonfourd\\
  \frac{\mixed}{\quarkthreed} \equiv M^{2}_{0} = (0.8\pm 0.1)\,\gev^2,
\end{gather}
where $\quarkthreed$ is given by PCAC using the coupling values
\begin{equation}
f_{\pi}=92.2\pm3.5\ \mev\ ,\
  f_K=110.0\pm4.2\ \mev
  .
\end{equation}
See Ref. \cite{HoHarnettSteele2016} for references to the sources for these parameters and uncertainties.
\begin{table}
\centering
\caption{Conventional meson data used to anchor double narrow resonance mixing analysis. Listed are lowest-lying conventional states with appropriate quantum numbers according to PDG \cite{OliveEtAl2014}. Entries have been omitted where no conventional meson state has been tabulated.
} 
\label{mixing_error}
\begin{tabular}{cccccc}
    Flavour & $J^{P}$ & PDG State & $m_{conv} (\gev)$ \\ 
\hline
 $\bar{c}Gq$ & $0^{+}$ &  $D^{*}_{0}\left(2400\right)^0$ & $2.318$ \\
  				  & $0^{-}$ &  $D^{0}$ & $1.865$\\
  				  & $1^{-}$ &  $D^{*}\left(2007\right)^0$ & $2.007$\\
    			  & $1^{+}$ &  $D_{1}\left(2420\right)^0$ & $2.420$\\
 $\bar{c}Gs$    & $0^{+}$ &  $D^{*}_{s0}\left(2317\right)^{\pm}$ & $2.318$\\
  				  & $0^{-}$ &  $D^{\pm}_{s}$ & $1.969$\\
  				  & $1^{-}$ &  $D^{*\pm}_{s}$ & $2.112$\\
    			  & $1^{+}$ &  $D_{s1}\left(2460\right)^{\pm}$ & $2.460$ \\
 $\bar{b}Gq$ & $0^{+}$ & - & - \\
  				  & $0^{-}$ &  $B_{0}$ & $5.279$ \\
  				  & $1^{-}$ &  $B^{*}$ & $5.324$ \\
    			  & $1^{+}$ &  $B_{1}\left(5721\right)^{0}$ & $5.726$\\
 $\bar{b}Gs$   & $0^{+}$ &  - & - \\
  				  & $0^{-}$ &  $B^{0}_{s}$ & $5.367$\\
  				  & $1^{-}$ &  $B^{*}_{s}$ & $5.416$\\
    			  & $1^{+}$ &  $B_{s1}\left(5830\right)^0$ & $5.828$
\end{tabular}
\end{table}
\section{Results of Laplace Sum-Rules Analysis}
\begin{figure}[!ht]
\centering
\includegraphics[width=\linewidth]{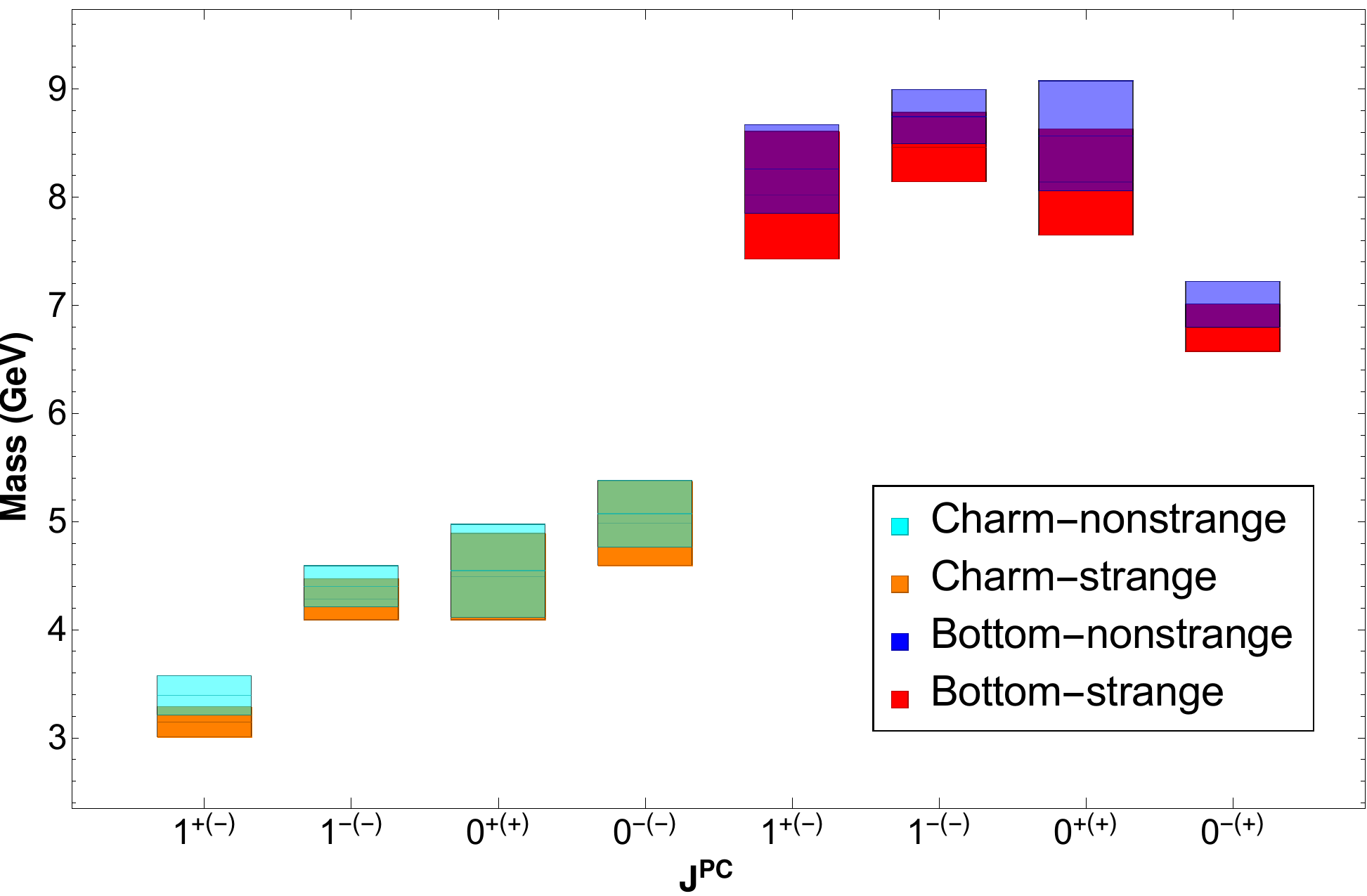}
\captionof{figure}{Summary of mass predictions with uncertainties for charm and bottom hybrid systems for the stabilizing $J^{P(C)}$ channels; channels which have been omitted do not stabilize.}\label{fig.MassSpectra}
\end{figure}

Performing a Laplace sum-rules analysis
of all eight distinct $J^{P(C)}$ combinations defined according to Table~\ref{JPC_table}, we present the mass predictions and estimated uncertainties in Figure \ref{fig.MassSpectra} and in Tables \ref{D0_results_table}-\ref{Bs_results_table}.  For the eight channels described in Table \ref{JPC_table}, only four stabilized for each combination of flavours ($\bar{c}Gq$, $\bar{c}Gs$, $\bar{b}Gq$, $\bar{b}Gs$). See Ref.~\cite{HoHarnettSteele2016} for an in-depth discussion and all results (including Borel windows and continuum parameters). A full uncertainty analysis was performed accounting for variations in QCD parameters (condensate values, heavy quark masses, mass ratios, $\alpha_s$ reference values, truncation of the OPE, and variations in the Borel window and renormalization scale) with contributions added in quadrature. Within the computed uncertainty, mass degeneracy between strange and non-strange channels cannot be precluded. 
\begin{table}[t]
\centering
\caption{Comparison of central values against GRW mass predictions for $\overline{c}qG$ hybrids ($q=\{u,d\}$).} 
\label{GRW_CharmCompare}
\begin{tabular}{cccccc}
    $J^{P}$ & $m_{\mathrm{GRW}}(\gev)$ & $m_{\mathrm{H}}\ (\gev)$ \\ 
\hline
  $0^{+}$ & $4.0$ &  $ 4.54$ \\
  $0^{-}$ & $4.5$ &  $ 5.07$  \\
  $1^{-}$ & $3.6$ & $ 4.40$  \\
  $1^{+}$ & $3.4$ & $ 3.39$  \\
\end{tabular}
\end{table}
\begin{table}
\centering
\caption{Comparison of central values against GRW mass predictions for $\overline{b}qG$ hybrids ($q=\{u,d\}$).}
\label{GRW_BottomCompare}
\begin{tabular}{cccccc}
    $J^{P}$ & $m_{\mathrm{GRW}}(\gev)$ & $m_{\mathrm{H}}\ (\gev)$ \\ 
\hline
  $0^{+}$ & $6.8$ & $8.57$ \\
  $0^{-}$ & $7.7$ & $7.01$   \\
  $1^{-}$ & $6.7$ & $8.74$  \\
  $1^{+}$ & $6.5$ & $8.26$  \\
\end{tabular}
\end{table}
Comparisons between our results and those found previously by GRW are shown in Tables \ref{GRW_CharmCompare} and \ref{GRW_BottomCompare}, where we report $J^P$ numbers because a change in channel stability was observed in our analysis; in GRW, the stable channels were
$J^{P(C)}\in\{0^{+(+)},0^{-(-)},1^{+(+)},1^{-(-)}\}$ for all heavy-light flavour hybrids. We see in Tables \ref{GRW_CharmCompare} and \ref{GRW_BottomCompare} that the central values of our predictions differ significantly from GRW, except in the case of $1^{+}$ charm-nonstrange. Additionally, there emerge similar mass hierarchies between the charm and bottom channels; excluding the $0^-$ states, the $1^+$, $1^-$ and $0^+$ states form a pattern where the $1^{+}$ state is lighter than essentially degenerate $1^{-}$ and $0^{+}$ states. While discrepancies in a shared hierarchy seem to appear with the inclusion of the $0^-$ states, we note that the charm and bottom $0^-$ mass predictions arise from interpolating currents with different C-parity. Although open-flavour systems have no well-defined $C$ quantum number,  Ref.~\cite{HilgerKrassnigg} attributes physical meaning to $C$ in the internal structures of hybrids; they find that the $0^{-(-)}$  structure is heavier than the $0^{-(+)}$, consistent with the pattern we observe in Fig.~\ref{fig.MassSpectra}. 

\begin{table}[!ht]
\centering
\caption{QCD sum-rules analysis results for ground state charm-nonstrange hybrids, including effect on hybrid mass prediction from mixing with conventional meson states; $\delta m_{mix}$ is the increased mass range with mixing up to $\left\vert\frac{f_{conv}}{f_H}\right\vert = \frac{1}{2}$ due to coupling to the lowest-lying conventional state with appropriate quantum numbers according to PDG values summarized in Table \ref{mixing_error}. }
\label{D0_results_table}
\begin{tabular}{cccccc}
    $J^{PC}$ & $m_H\pm\delta m_{H}\ (\gev)$ & $f_{H}^{2}\times10^{6}$&  $+\delta m_{mix}~(\gev)$ \\ 
\hline
  $0^{+(+)}$ & $ 4.55\pm 0.43$ & $7.47$ & $0.02$\\
  $0^{-(-)}$ & $ 5.07\pm 0.31$ & $7.28$ & $0.00$\\
  $1^{-(-)}$ & $ 4.40\pm 0.19$ & $12.4$ & $0.01$\\
  $1^{+(-)}$ & $ 3.39\pm 0.18$ & $9.87$ & $0.05$
\end{tabular}
\end{table}
\begin{table}[!ht]
\centering
\caption{QCD sum-rules analysis results for ground state charm-strange hybrids, including effect on hybrid mass prediction from mixing with conventional meson states; $\delta m_{mix}$ is the increased mass range with mixing up to $\left\vert\frac{f_{conv}}{f_H}\right\vert = \frac{1}{2}$ due to coupling to the lowest-lying conventional state with appropriate quantum numbers according to PDG values summarized in Table \ref{mixing_error}. }
\label{Ds_results_table}
\begin{tabular}{cccccc}
    $J^{PC}$ & $m_H\pm\delta m_{H}\ (\gev)$ & $f_{H}^{2}\times10^{6}$&  $+\delta m_{mix}~(\gev)$ \\ 
\hline
  $0^{+(+)}$ & $ 4.49\pm 0.40$ & $7.36 $ & $0.02$ \\
  $0^{-(-)}$ & $ 4.98\pm 0.39$ & $2.03 $ & $0.00$ \\
  $1^{-(-)}$ & $ 4.28\pm 0.19$ & $11.0 $ & $0.02$ \\
  $1^{+(-)}$ & $ 3.15\pm 0.14$ & $8.45 $ & $0.06$ 
\end{tabular}
\end{table}
\begin{table}[!ht]
\centering
\caption{QCD sum-rules analysis results for ground state bottom-nonstrange hybrids, including effect on hybrid mass prediction from mixing with conventional meson states; $\delta m_{mix}$ is the increased mass range with mixing up to $\left\vert\frac{f_{conv}}{f_H}\right\vert = \frac{1}{2}$ due to coupling to the lowest-lying conventional state with appropriate quantum numbers according to PDG values summarized in Table \ref{mixing_error}.  }
\label{B0_results_table}
\begin{tabular}{cccccc}
    $J^{PC}$ & $m_H\pm\delta m_{H}\ (\gev)$& $f_{H}^{2}\times10^{6}$&  $+\delta m_{mix}~(\gev)$ \\ 
\hline
  $0^{+(+)}$ & $8.57\pm 0.51$ & $1.28$ & - \\
  $0^{-(+)}$ & $7.01 \pm 0.21$ &  $0.516$ & $0.19$\\
  $1^{-(-)}$ & $8.74\pm 0.25$ & $1.76$ & $0.32$\\
  $1^{+(-)}$ & $8.26\pm 0.41$ & $1.66$ & $0.74$
\end{tabular}
\end{table}
\begin{table}
\centering
\caption{QCD sum-rules analysis results for ground state bottom-strange hybrids, including effect on hybrid mass prediction from mixing with conventional meson states; $\delta m_{mix}$ is the increased mass range with mixing up to $\left\vert\frac{f_{conv}}{f_H}\right\vert = \frac{1}{2}$ due to coupling to the lowest-lying conventional state with appropriate quantum numbers according to PDG values summarized in Table \ref{mixing_error}.}
\label{Bs_results_table}
\begin{tabular}{cccccc}
    $J^{PC}$ & $m_H\pm\delta m_{H}\ (\gev)$& $f_{H}^{2}\times10^{6}$&  $+\delta m_{mix}~(\gev)$ \\ 
\hline
  $0^{+(+)}$ & $8.14\pm 0.49$ &$0.817$ & -\\
  $0^{-(+)}$ & $6.79\pm 0.22$ &$0.434$ & $0.44$\\
  $1^{-(-)}$ & $8.46\pm 0.32$ &$1.24$  & $0.35$\\
  $1^{+(-)}$ & $8.02\pm 0.59$ &$1.39 $ & $0.72$
\end{tabular}
\end{table}
\section{Conclusions}
Ground state mass predictions of heavy-light (strange and nonstrange) hybrid mesons for $J^{P} \in \left\{ 0^{\pm},~1^{\pm}\right\}$ have been briefly presented, utilizing QCD sum-rules and improving upon previous calculations \cite{GovaertsReindersWeyers1985} by updating the non-perturbative parameters in the calculation, and including higher dimensional condensates in the OPE shown important to sum-rule stability in other contexts. A complete discussion of the analysis and results may be found in \cite{HoHarnettSteele2016}. A degeneracy is observed in the heavy-light and heavy-strange states, 
and stabilization in the previously unstable $0^{-(-)}$ and $1^{-(+)}$ channels \cite{GovaertsReindersWeyers1985} driven by the addition of the higher dimensional 5d mixed and 6d gluon condensate contributions. As a consequence of these higher dimensional contributions, the $1^{+(+)}$ channel is destabilized from the original analysis of \cite{GovaertsReindersWeyers1985}.
Possible mixing effects are examined, and in our simplest mixing model we find our predictions serve as lower bounds on ground state mass predictions. 

\textbf{Acknowledgements-} We are grateful for financial support from the Natural Sciences and Engineering Research Council of Canada (NSERC).
\bibliographystyle{elsarticle-num} 
\bibliography{ichep2016}
\end{document}